\documentstyle[preprint,revtex]{aps}
\begin{document}
\begin{title}
Rotational covariance and light-front current matrix elements
\end{title}
\author{B. D. Keister}
\begin{instit}
Department of Physics, Carnegie Mellon University, Pittsburgh, PA 15213
\end{instit}
\begin{abstract}
Light-front current matrix elements for elastic scattering from
hadrons with spin~1 or greater must satisfy a nontrivial constraint
associated with the requirement of rotational covariance for the
current operator.  Using a model $\rho$ meson as a prototype for
hadronic quark models, this constraint and its implications are
studied at both low and high momentum transfers.  In the kinematic
region appropriate for asymptotic QCD, helicity rules, together with
the rotational covariance condition, yield an additional relation
between the light-front current matrix elements.
\end{abstract}
\newpage
\narrowtext
\section{Introduction}
Light-front dynamics has found frequent application in particle and
nuclear physics.  First introduced by Dirac~\cite{Dirac}, it has the
advantage that, of the ten generators of transformations for the
Poincar\'e group, only three of them depend upon interaction dynamics.
In particular, certain forms of Lorentz boosts are interaction
independent.  This is in contrast to the more traditionally used
instant-form dynamics, which has four interacting generators,
including {\it all} boosts.

However, as with all forms of relativistic dynamics, more generators
than just the light-front ``Hamiltonian'' $P^- = P^0 - {\bf
P}\cdot{\bf n}$, where ${\bf n}$ is a spatial unit vector, must depend
nontrivially on the interaction, and these generators correspond to
rotations about axes perpendicular to ${\bf n}$.  This is especially
important when computing matrix elements of electromagnetic and weak
currents in the front form.  Because the current operator must have
the transformation properties of a four-vector, and some of these
transformations are interaction dependent, it must also depend upon
the strong interaction.  In particular, the current operator cannot
satisfy all of the covariance requirements associated with transverse
rotations without containing interaction dependent components.  This
is sometimes called the ``angular condition''~\cite{GellMann,Osborn}.

Rotational covariance has been studied in a variety of contexts.  For
hadrons composed of quarks, Terent'ev and others have examined the
extent to which the current operator is uniquely determined without
knowing its two-body components.  For elastic electron-deuteron
scattering, the lack of uniqueness modulo two-body currents has been
explored by employing a variety of ``schemes'' to satisfy the
rotational covariance requirement~\cite{Grach,CCKP}.

In this paper, we examine a specific rotational constraint for
light-front current matrix elements of hadrons composed of quarks.
The current operator must satisfy certain nontrivial commutation
relations with the interacting generators of the Poincar\'e group.
The requirement of current covariance for rotations about an axis
perpendicular to ${\bf n}$ gives rise to constraints on current matrix
elements for elastic scattering from particles of spin 1 or
greater~\cite{Grach,CCKP}.  While form factors of the pion~\cite{CCP}
and the nucleon~\cite{CC} have frequently been calculated, the
particular test discussed here is not applicable because $j < 1$.  We
therefore take a model of a $\rho$ meson as a prototype hadron to
which to apply the rotational covariance test.
\section{Rotational Covariance}
In light-front dynamics, full rotational covariance implies a
nontrivial set of conditions which any hadronic model must in
principle satisfy.

First, the state vector for the hadron must be an eigenstate of the
total spin operator.  This condition is satisfied in light-front
models which use a quantum mechanical Hilbert space with a fixed
number of particles~\cite{CCP,CC}.  Models using a field theory, in
particular those motivated by QCD, may not necessarily use
rotationally covariant state vectors.  At high $Q^2$, this deficiency
may be irrelevant if the corrections to rotational covariance fall as
a power of $Q^2$.  At moderate and low $Q^2$, the issue may be
important.  In any event, we consider here only models whose state
vectors have the proper rotational covariance properties.

Second, the current operator $I^\mu(x)$ must satisfy the conditions of
Lorentz covariance.  If $\Lambda^{\mu\nu}$ is the matrix for a
homogeneous Lorentz transformation and $a^\mu$ is a spacetime
translation, then
\begin{equation}
U({\Lambda} ,{a})I^{\mu} (x) U ({\Lambda} ,{a})^{\dag} =
(\Lambda^{-1})^{\mu}{}_{\nu} I^{\nu} (\Lambda x+a).
\label{kAD}
\end{equation}
It must be conserved with respect to the four-momentum. If $P^\mu$ is
the generator of spacetime translations, then
\begin{equation}
g_{\mu \nu}[P^{\mu} , I^{\nu} (0) ] = 0.
\label{kAE}
\end{equation}
These constraints have two implications.  First, it is possible to
express the physical content of current matrix elements between any
two states in terms of a limited number of Lorentz invariant functions
of the masses of the states and the square of the momentum transfer.
Second, the operator $I^\mu(0)$ must have in general a complicated
structure, since it obeys nontrivial commutation relations with at
least some generators which are interaction dependent.

To illustrate these two points, consider current matrix elements with
spacelike momentum transfer.  It has been shown~\cite{CCKP,KP} that
all spin matrix elements of the current four-vector can be computed
from the set of matrix elements of $I^+(0)$ in a frame in which
$q^+ = q^0 + {\bf q}\cdot{\bf n} = 0$.  Alternatively, it means
that all invariant form factors can be computed from matrix elements
of $I^+(0)$.   However, the covariance requirements imply that the
matrix elements of $I^+(0)$ must satisfy a set of constraints.  This
is particularly relevant if a one-body operator has been used to
compute the matrix elements.  If a constraint involves
transformations which use Poincar\'e generators which are
non-interacting, it will in general be satisfied for matrix
elements computed with one-body operators.  However, constraints
which involve transformations using interacting generators will in
general not be satisfied with one-body current matrix elements.

An interaction-dependent constraint can be derived by requiring that
Breit-frame matrix elements of the transverse current in a helicity
basis vanish if the magnitude of the helicity flip is 2 or greater.
For light-front current matrix elements $\langle {\tilde{\bf p}}'\mu'
| I^+(0) | {\tilde{\bf p}}\mu\rangle $ corresponding to elastic
electron scattering from a target of mass $M$ and spin $j$, the
condition is expressed as follows:
\begin{equation}
\sum_{\lambda'\lambda}
D^{(j){\dag}}_{\mu'\lambda'}(R'_{ch})
\langle {\tilde{\bf p}}'\alpha' | I^+(0) | {\tilde{\bf p}}\alpha\rangle
D^{(j)}_{\lambda\mu}(R_{ch}) = 0,\quad |\mu'-\mu| \ge 2.
\label{BAA}
\end{equation}
In Eq.~(\ref{BAA}), ${\tilde{\bf p}}\equiv({\bf p}_\perp,p^+)$ is a
light-front momentum, and the matrix element is evaluated in a frame
where $q^+ = p'{}^+ - p^+ = 0$, and the perpendicular component of
${\tilde{\bf p}}'$ and ${\tilde{\bf p}}$ lies along the $x$ axis.  The
rotation
\begin{equation}
R_{ch} = R_{cf}({\tilde{\bf p}}, M) R_y({\textstyle{\pi\over2}}); \quad
R'_{ch} = R_{cf}({\tilde{\bf p}}', M) R_y({\textstyle{\pi\over2}}),
\label{BAABA}
\end{equation}
where $R_{cf}$ is a Melosh rotation which, together with the rotation
$R_y({\pi\over2})$, transforms the state vectors from light-front spin
to helicity.  For inelastic excitation of a state with mass $M'$ and
spin $j'$, Eq.~(\ref{BAA}) is modified only by the use of $M'$ and
$j'$ in the rotation matrices for the final state.  For elastic
scattering, Eq.~(\ref{BAA}) is applicable only to states with $j\ge
1$.  For a spin-1 particle, Eq.~(\ref{BAA}) can be expressed
explicitly in terms of individual light-front matrix elements as
follows~\cite{Grach}:
\begin{equation}
\Delta(Q^2) \equiv (1 + 2 \eta) I_{1,1} + I_{1,-1}
- \sqrt{8\eta} I_{1,0} - I_{0,0} = 0,
\label{BBB}
\end{equation}
where $I_{\mu',\mu} \equiv
\langle {\tilde{\bf p}}' \mu' | I^+(0) | {\tilde{\bf p}} \mu\rangle $
is the matrix element of $I^+(0)$:
\begin{equation}
{\bf p}_\perp' = - {\bf p}_\perp = {\textstyle{1\over2}} {\bf q};\quad
p'{}^+ = p^+ = \sqrt{M^2 + {\textstyle{1\over4}}{\bf q}^2},
\label{BBBA}
\end{equation}
$Q^2 = -q^2$ is the square of the four-momentum transfer, $\eta \equiv
Q^2 / 4 M^2$.

Earlier studies of the deuteron form factors using models with
one-body currents and rotationally covariant state vectors indicate
that $\Delta(Q^2)$ can be relatively small for low and moderate $Q^2$,
though, not surprisingly, $\Delta(Q^2)$ is an increasing function of
$Q^2$~\cite{Grach,CCKP}.  An important feature of the deuteron is that
the characteristic nucleon momentum is very small compared to a
nucleon mass.

For most quark models of hadrons, the characteristic constituent
momentum is not small with respect to the quark mass, and questions of
rotational covariance therefore require a separate investigation.
There have been previous studies of the pion~\cite{CCP} and the
nucleon~\cite{CC} form factors using models with one-body currents and
rotationally covariant state vectors.  Equation~(\ref{BBB}) has no
counterpart for spin zero and spin ${1\over2}$, so the dynamical
aspect of rotational covariance was not addressed in those works.
Some other conclusions from those studies which are relevant to the
discussion below include
\begin{enumerate}
\item For small quark masses (10 MeV), relativistic effects such as
those of Melosh rotations can be substantial, even at low $Q^2$.
\item In the limit $m_q\to 0$, it can be shown~\cite{Terentev} that
$Q^2 F_\pi(Q^2) \to {\rm const}$ as $Q^2\to\infty$.
\end{enumerate}
In what follows, we examine a model for a $\rho$ meson similar to
those used for the pion published earlier, in light of the rotational
covariance condition~(\ref{BBB}) at both low and high $Q^2$.
\section{The Model}
The model $\rho$ meson is composed of a valence quark and an
antiquark.  Since it must be color-antisymmetric and flavor-symmetric
(isospin 1), the space-spin wave function must be symmetric.  For the
ground state, we take the coupled states ${{}^3S_1-{}^3D_1}$.  This is
the same space-spin combination as that of a deuteron composed of two
nucleons.  Since the details of such a deuteron form factor
calculation are discussed extensively elsewhere~\cite{CCKP}, only the
unique features of the quark model will be presented here.  The mass
operator is
\begin{equation}
M^2 = M_0^2 + 2m U + U_0,
\label{BBBAC}
\end{equation}
where $M_0^2 = 4(m^2 + {\bf k}^2)$ is the non-interacting mass
operator, $m$ is the quark mass and ${\bf k}$ is the relative
three-momentum.  The potential $U$ is a harmonic-oscillator potential:
\begin{equation}
U = {\textstyle{1\over2}} \kappa {\bf r}^2,\quad
{\bf r} \equiv i {{\mbox{\boldmath $\nabla$}}}_{\bf k},
\label{ABJ}
\end{equation}
and $U_0$ is a constant
The $S$- and $D$-state wave functions are Gaussians, just as for a
non-relativistic harmonic-oscillator problem:
\begin{eqnarray}
\phi_0(k) && = N_S e^{-k^2 / 2 \alpha^2}; \nonumber  \\
\phi_2(k) && = N_D k^2 e^{-k^2 / 2 \alpha^2}.
\label{BBC}
\end{eqnarray}
The interacting mass eigenvalue is
\begin{equation}
M^2 = 4(m^2 + 3 \alpha^2) + U_0.
\label{ABG}
\end{equation}
In an extensive study of mesons using a nonrelativistic quark model,
Godfrey and Isgur obtained $D$-state admixtures with amplitude 0.04
for an excited $\rho$-meson state, with no reported admixture for the
$\rho$ (750)~\cite{Godfrey}.  To test sensitivity, we use a $D$-state
admixture amplitude coefficient of 0.04; this is an extreme choice,
but in the end, the results differ little from those obtained by
ignoring the $D$-state admixture entirely.  An oscillator parameter
value $\alpha = 0.45$ GeV/$c$ has been extracted from the
Godfrey-Isgur results~\cite{Kokoski}.
\section{Form Factors}
Matrix elements of matrix elements of $I^+(0)$ can be written
as~\cite{CCKP}
\begin{eqnarray}
&&\langle {\tilde{\bf p}}'\mu' | I^+(0) | {\tilde{\bf p}} \mu\rangle
\nonumber  \\
&&\qquad =  \langle p_1^+\,{\textstyle{1\over2}}{\bf q}_\perp \mu_1'|
I_{\rm quark}^+(0)|p_1^+\,-{\textstyle{1\over2}}{\bf q}_\perp \mu_1\rangle
\int{d{\bf k}\over{(2\pi)^3}}
\left|{{\partial({\tilde{\bf p}}{\bf k})}\over
\partial({\tilde{\bf p}}_1{\tilde{\bf p}}_2)}\right| ^{1\over2}
\left|{\partial({\tilde{\bf p}}_1'{\tilde{\bf p}}_2)\over
{\partial({\tilde{\bf p}}'{\bf k}')}}\right|
^{1\over2} \nonumber  \\
&&\qquad\quad\times
D^{({1\over2})}_{{\bar\mu_1} \mu_1}
\left[R_{cf}({\bf k}')\right]
D^{({1\over2})}_{{\bar\mu_2} \mu_2}
\left[R_{cf}(-{\bf k}')\right] \nonumber  \\
&&\qquad\quad\times
\langle {\textstyle{1\over2}} \bar\mu_1'
{\textstyle{1\over2}} \bar\mu_2' | 1 \mu_S' \rangle
\langle l \mu_l' 1 \mu_S' | 1 \mu'\rangle
Y_{l'\mu_l'} ({\bf\hat k}') \phi_{l'}(k') \nonumber  \\
&&\qquad\quad\times
D^{({1\over2}){\dag}}_{\mu_1 \bar \mu_1}
\left[R_{cf}({\bf k})\right]
D^{({1\over2}){\dag}}_{\mu_2 \bar \mu_2}
\left[R_{cf}({\bf k})\right] \nonumber  \\
&&\qquad\quad\times
\langle {\textstyle{1\over2}} \bar\mu_1
{\textstyle{1\over2}} \bar\mu_1 | 1 \mu_S \rangle
\langle l \mu_l 1 \mu_S | 1\mu\rangle
Y_{l\mu_l} ({\bf\hat k}) \phi_l(k).
\label{BBCA}
\end{eqnarray}
The internal kinematics in the integral are
\begin{eqnarray}
p_1'{}^+ &&= p_1^+ = \sqrt{m^2 +
{\textstyle{1\over4}}{\bf q}_\perp^2};
\qquad p_2'{}^+ = p_2^+;\qquad
{\bf p}'_{1\perp} = {\bf p}_{1\perp} + {\bf q}_\perp;
\qquad {\bf p}'_{2\perp} = {\bf p}_{2\perp} \nonumber  \\
{\bf k}_\perp &&= (1-\xi){\bf p}_{1\perp} - \xi {\bf p}_{2\perp};
\qquad {\bf k}'_\perp = {\bf k}_\perp + (1-\xi){\bf q}_\perp;
\qquad \xi \equiv p_1^+ / (p_1^+ + p_2^+) \nonumber  \\
k_3 &&= (\xi - {\textstyle{1\over2}})
\sqrt{m^2 + {\bf k}_\perp^2 \over \xi (1-\xi)};\qquad
k'{}_3 = (\xi - {\textstyle{1\over2}})
\sqrt{m^2 + {\bf k}'_\perp{}^2 \over \xi (1-\xi)}; \nonumber  \\
{\bf k} &&= ({\bf k}_\perp,k_3);\qquad
{\bf k}' = ({\bf k}'_\perp,k'_3).
\label{BBCAAA}
\end{eqnarray}

The three elastic form factors $G_1$, $G_2$ and $G_3$ can be expressed
in terms of the matrix elements $I_{\mu',\mu}$ as
follows~\cite{CCKP}:
\begin{eqnarray}
G_0(Q^2) &&= {1\over 2(1+\eta)}
\left[(1-{\textstyle{2\over3}}\eta)(I_{1,1}+I_{0,0})
+{\textstyle{5\over3}}\sqrt{8\eta} I_{1,0}
-{\textstyle{1\over3}}(1-4\eta)I_{1,-1}\right] \nonumber  \\
G_1(Q^2) &&= {1\over (1+\eta)}
\left[I_{1,1} + I_{0,0} - I_{1,-1}
- (1-\eta)\sqrt{2\over\eta} I_{1,0}\right] \nonumber  \\
G_2(Q^2) &&= {\sqrt{8}\over 3(1+\eta)}
\left[-{\eta\over2}(I_{1,1}+I_{0,0})
+\sqrt{2\eta} I_{1,0}
-(1+{\textstyle{1\over2}}\eta)I_{1,-1}\right].
\label{BBD}
\end{eqnarray}
The right-hand sides in Eq.~(\ref{BBD}) are not unique.  One can always
replace one of the $I_{\mu',\mu}$, or linear combinations of
them, with a combination which satisfies the rotational covariance
condition~(\ref{BBB}).  A common procedure has been to choose a
particular combination of $I_{\mu',\mu}$ {\it as calculated
from one-body current operators,} and eliminating the remaining terms
from Eq.~(\ref{BBD}) via the rotational covariance condition.
By implication, the
eliminated terms depend upon two-body current operators.  Thus, for
different choices of one-body matrix elements, each form factors $G_i(Q^2)$
will differ by a multiple of $\Delta(Q^2)_{\rm one-body}$, which is
never zero.
\section{Low $Q^2$}
The requirement of rotational covariance can be studied at low
momentum transfers by examining the behavior of the magnetic and
quadrupole moments $\mu$ and $\bar Q$ and the charge radius.  They are
related to the form factors $G_i(Q^2)$ as follows:
\begin{eqnarray}
\mu &&\equiv \lim_{Q^2 \to 0} G_1(Q^2) \nonumber  \\
{\bar Q} &&\equiv  \lim_{Q^2 \to 0} 3\sqrt{2} {G_2(Q^2)\over Q^2}
\nonumber  \\
\langle r^2\rangle  &&\equiv \lim_{Q^2 \to 0} {6\over Q^2} [1 - G_0(Q^2)].
\label{BBE}
\end{eqnarray}
Extracted values of $\langle r^2\rangle ^{1\over2}$ are shown in
Table~\ref{rsq} for quark masses of 10, 300 and 1000 MeV.  Also shown
is the effect of including or leaving out the Melosh rotations, which
gives a measure of the size of relativistic effects.  In addition, the
quantity
\begin{equation}
\delta \equiv \lim_{Q^2\to 0} {\Delta(Q^2)\over 1-G_0(Q^2)}
\label{BBEA}
\end{equation}
gives a measure of the sensitivity to rotational covariance
uncertainty.  Already one can see from this table corresponding to
very low $Q^2$ that simply raising the value of the quark mass is not
the same as the nonrelativistic limit.  That limit depends upon the
quark mass, the value of $Q^2$, the momentum scale $\alpha$, {\it and
the composite mass.} For comparison, we also show results using the
same parameter, except that the $\rho$ meson is given a fictitious
value of 2 GeV.  In this last case, especially for a quark mass of
1~GeV, one can see that both the Melosh rotations (the measure of
relativistic effects) and the rotational covariance parameter $\delta$
are small.
\section{Moderate $Q^2$}
In Figs.~\ref{Ga750}, \ref{Gb750} and~\ref{Gc750} we show the
calculated results for $G_0(Q^2)$, as obtained using Eq.~(\ref{BBD}).
The relativistic effects, as characterized by turning the internal
Melosh rotations on and off, are largest for the smallest quark mass.
For all three choices of quark mass, the rotational covariance
uncertainty function $\Delta(Q^2)$ becomes comparable to $G_0$, and
hence the current matrix elements themselves, in the region 1--2 ${\rm
GeV}/c^2$.

In Figs.~\ref{Ga2000}, \ref{Gb2000} and~\ref{Gc2000} we show
calculated results for $G_0(Q^2)$ using a fictitious $\rho$-meson mass
of 2 GeV.  In all three cases, the relativistic effects are smaller
than the corresponding cases where $M_\rho$ = 750 MeV, but the
nonrelativistic limit is still not really reached until the quark
masse is considerably larger than the momentum scale $\alpha$.  For
this choice of meson mass, and for all three choices of quark mass,
the covariance function $\Delta(Q^2)$ is much smaller for the same
range of $Q^2$ than in the previous three figures.

{}From the results shown here, along with those of other
parameter sets, it becomes clear that the rotational covariance
uncertainty function $\Delta(Q^2)$ becomes comparable to $G_0$, and
hence the current matrix elements themselves, when ${\eta = Q^2 / 4M^2}$
is of order unity.  Thus, for a $\rho$ meson with physical mass 750
MeV, the breakdown of rotational covariance occurs in the region 1--2
${\rm GeV}/c^2$.  This can be understood from the fact that the
dimensionless argument of the Melosh rotations in Eq.~(\ref{BAA}) is
$Q/2M$, which manifests itself in terms of the $\eta$ factors in
Eq.~(\ref{BBB}).  The dynamical nature of the rotational covariance
condition is contained in the presence of the interacting mass $M$.

Note also that, for elastic scattering, the current matrix elements
$I_{\mu',\mu}$ depend upon the quark mass $m$ and the momentum
transfer, {\it but they do not depend upon the composite mass $M$}.
The internal Melosh rotations, which give a measure of size of
relativistic effects, depend upon the quark mass, but the composite
mass enters {\it only} at the point of computing form factors and
evaluating the rotational covariance condition.
\section{Asymptotic Behavior}
As noted above, it has been shown that, for the pion form factor,
models such as the one used here have the property that
$Q^2F_\pi(Q^2)\to {\rm const}$ as $Q^2\to\infty$ if $m_q=0$.  For a
model $\rho$ meson, the differing feature is the presence of an
overall spin index and some momentum-independent Clebsch-Gordan
coefficients.  We therefore expect that, as for the case of the pion
form factor~\cite{Terentev}, $Q^2 I_{\mu',\mu} \to {\rm const}$
(perhaps dependent upon ${\mu',\mu}$) as $Q^2\to\infty$ for $m_q=0$
and any $\mu',\mu$.  On the other hand, power counting rules of
perturbative QCD~\cite{CarlsonGross} predict that the matrix element
$I_{0,0}$ dominates as $Q^2\to\infty$, and that $I_{1,0}$ is
suppressed by on power of $Q$ and $I_{1,-1}$ by two powers of $Q$.
Thus, a constituent quark model with one-body currents only cannot
reproduce the asymptotic limit.

While simple models may fail to describe the asymptotic limit
appropriate for perturbative QCD, we note that the rotational
covariance condition
takes a simple form at very high $Q^2$.  In the $Q^2\to\infty$ limit
of Eq.~(\ref{BBB}), $I_{1,-1}$ drops out due to power suppression.
The remaining terms give
\begin{equation}
2 \eta I_{1,1} - \sqrt{8\eta} I_{1,0} - I_{0,0} = 0,
\label{BBF}
\end{equation}
The Breit frame $(1,1)$ matrix element of the transverse current in a
helicity basis is identically zero~\cite{CarlsonGross}.  The
light-front matrix element $I_{1,1}$ is not identically zero, but is
suppressed by two powers of $Q$ relative to a specific combination of
$I_{0,0}$ and $Q I_{1,0}$.

Note also that, in Eq.~(\ref{BBF}), the factors of $\eta$ which
correspond to the dynamical nature of the rotational covariance
condition are now very large.  At the same time, the gluon-exchange
terms used to derive the power-counting helicity rules in perturbative
QCD correspond to two-body currents in a constituent model such as
that presented here.  At high $Q^2$, therefore, the $\eta$ factors,
the dynamics of perturbative QCD and rotational covariance are linked
in a way which cannot be described in a constituent model with
one-body currents.
\section{Conclusions}
The requirement of rotational covariance for matrix elements of
electromagnetic currents is nontrivial to satisfy for for elastic
scattering from systems with spin $j\ge 1$.  For the case of the
deuteron, the fact that its structure is essentially nonrelativistic
(all masses are large compared to the momentum scale of the system)
suggests that the rotational covariance requirement can be satisfied
in a satisfactory quantitative way using only one-body current matrix
elements.  For hadrons composed of quarks, typical quark masses are
not small compared to typical momentum scales, and issue of rotational
covariance therefore must be studied separately.  In this paper, we
have investigated the behavior of current matrix elements in a simple
model of a $\rho$~meson.  Our results using variable input parameters
indicate a breakdown of rotational covariance of current matrix
elements when ${\eta = Q^2 / 4M^2}$ is of order unity.  The dynamical
nature of the rotational covariance constraint is reflected in the
presence of the interacting mass eigenvalue of the composite particle.
In addition, rotational covariance implies a specific power-law
relation among the spin matrix elements at high $Q^2$.  That relation
is consistent with the power-counting rules of perturbative QCD, which
in turn are derived from gluon-exchange contributions that correspond
to two-body currents in a constituent-quark framework.  The quark
model used in this paper does not contain such two-body currents, and
also does not satisfy the high-$Q^2$ power-law relation.
%
\section{Acknowledgements}
This work was supported in part by the U.S. National Science
Foundation under Grant PHY-9023586.  The author wishes to thank
Professor C. E. Carlson, Dr.~P.-L. Chung and Professor N. Isgur for
helpful conversations and correspondence.
\mediumtext
\newif\ifmultiplepapers
\def\beginpapers{\multiplepaperstrue}
\def\endpapers{\multiplepapersfalse}
\def\journal#1&#2(#3)#4{\sl #1~{\bf #2}\unskip, \rm  #4 (19#3)}
\def\trjrnl#1&#2(#3)#4{\sl #1~{\bf #2} \rm #4 (19#3)}
\def\baps{\journal {Bull.} {Am.} {Phys.} {Soc.}&}
\def\jap{\journal J. {Appl.} {Phys.}&}
\def\prl{\journal {Phys.} {Rev.} {Lett.}&}
\def\pl{\journal {Phys.} {Lett.}&}
\def\pr{\journal {Phys.} {Rev.}&}
\def\np{\journal {Nucl.} {Phys.}&}
\def\rmp{\journal {Rev.} {Mod.} {Phys.}&}
\def\jmp{\journal J. {Math.} {Phys.}&}
\def\rmm{\journal {Revs.} {Mod.} {Math.}&}
\def\jetp{\journal {J.} {Exp.} {Theor.} {Phys.}&}
\def\sjetp{\trjrnl {Sov.} {Phys.} {JETP}&}
\def\dokl{\journal {Dokl.} {Akad.} Nauk USSR&}
\def\spd{\trjrnl {Sov.} {Phys.} {Dokl.}&}
\def\tmf{\journal {Theor.} {Mat.} {Fiz.}&}
\def\snp{\trjrnl {Sov.} J. {Nucl.} {Phys.}&}
\def\hpa{\journal {Helv.} {Phys.} Acta&}
\def\yf{\journal {Yad.} {Fiz.}&}
\def\zp{\journal Z. {Phys.}&}
\def\anp{\journal {Adv.} {Nucl.} {Phys.}&}
\def\ap{\journal {Ann.} {Phys.}&}
\def\am{\journal {Ann.} {Math.}&}
\def\nc{\journal {Nuo.} {Cim.}&}
\def\pre{\journal {Phys.} {Rep.}&}
\def\pca{\journal Physica (Utrecht)&}
\def\prs{\journal {Proc.} R. {Soc.} London &}
\def\jcp{\journal J. {Comp.} {Phys.}&}
\def\pna{\journal {Proc.} {Nat.} {Acad.}&}
\def\jpg{\journal J. {Phys.} G (Nuclear Physics)&}
\def\fort{\journal {Fortsch.} {Phys.}&}
\def\jfa{\journal {J.} {Func.} {Anal.}&}
\def\cmp{\journal {Comm.} {Math.} {Phys.}&}

%
\pagebreak
\begin{table}
\caption{Calculated r.m.s.\ charge radius, together with the
dimensionless rotational covariance parameter $\delta$, for various
model parameters.}
\label{rsq}
\begin{tabular}{cccrr}
\multicolumn{1}{c} {$M_\rho$ (MeV)} &
\multicolumn{1}{c} {$m_q$ (MeV)} & \multicolumn{1}{c} {Melosh} &
\multicolumn{1}{c} {$\langle r^2\rangle ^{1\over2}$ (fm)} &
\multicolumn{1}{c} {$\delta$} \\
\tableline
750	& 10	& ON	& 1.17	&  .03	\\
	&	& OFF	& 0.76	&  .16	\\
	& 300	& ON	& 0.61	&  .17	\\
	&	& OFF	& 0.54	&  .36	\\
	& 1000	& ON	& 0.51	&  .36	\\
	&	& OFF	& 0.50	&  .42	\\
2000	& 10	& ON	& 1.11	&  .11	\\
	&	& OFF	& 0.66	&  .05	\\
	& 300	& ON	& 0.45	&  .19	\\
	&	& OF	& 0.37	&  .11	\\
	& 1000	& ON	& 0.33	&  .04	\\
	&	& OFF	& 0.32	&  .15	\\
\end{tabular}
\end{table}
\mediumtext
\figure{%
Composite form factor $G_0(Q^2)$ computed using a quark mass of 10 MeV
and a $\rho$-meson mass of 750 MeV.  The solid curve denotes the full
calculation and the dashed curve corresponds to the calculation where
the Melosh rotations are turned off.  The dot-dashed curve describes
the covariance function $\Delta(Q^2)$ defined in Eq.~(\ref{BBB}).
\label{Ga750}}
\figure{%
Composite form factor $G_0(Q^2)$ computed using a quark mass of 300 MeV
and a $\rho$-meson mass of 750 MeV.  The legend is the same as that of
Fig.~\ref{Ga750}.
\label{Gb750}}
\figure{%
Composite form factor $G_0(Q^2)$ computed using a quark mass of 1 GeV
and a $\rho$-meson mass of 750 MeV.  The legend is the same as that of
Fig.~\ref{Ga750}.
\label{Gc750}}
\figure{%
Composite form factor $G_0(Q^2)$ computed using a quark mass of 10 MeV
and a $\rho$-meson mass of 2 GeV.  The legend is the same as that of
Fig.~\ref{Ga750}.
\label{Ga2000}}
\figure{%
Composite form factor $G_0(Q^2)$ computed using a quark mass of 300 MeV
and a $\rho$-meson mass of 2 GeV.  The legend is the same as that of
Fig.~\ref{Ga750}.
\label{Gb2000}}
\figure{%
Composite form factor $G_0(Q^2)$ computed using a quark mass of 1 GeV
and a $\rho$-meson mass of 2 GeV.  The legend is the same as that of
Fig.~\ref{Ga750}.
\label{Gc2000}}

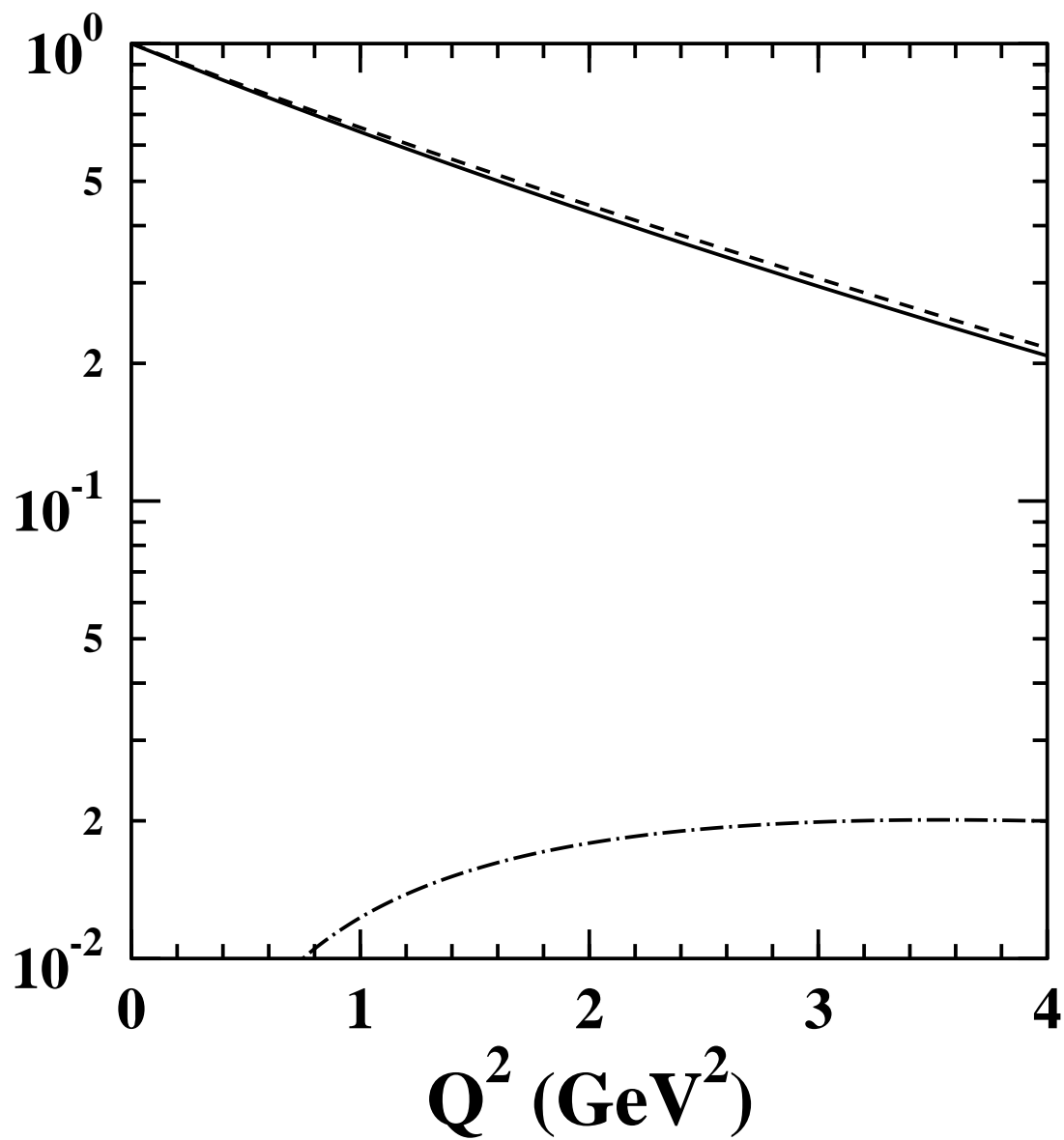
\end{document}